\documentclass[a4paper,11pt]{article}
\pdfoutput=1 
\usepackage{jheppub}

\title{\boldmath Are Leptons as elementary as Quarks?}

\author{Aharon Davidson}
\affiliation{Physics Department, Ben-Gurion University
of the Negev, \\ Beer-Sheva 84105, Israel}

\emailAdd{davidson@bgu.ac.il}

\abstract{Nucleons and electrons were once
	considered elementary particles, a role
	nowadays taken by quarks and leptons.
	Here, mainly at the group theoretical level,
	we examine the unorthodox idea that
	nucleons and electrons share the same
	level of compositeness after all.
	We do it by first trading color $SU(3)_C$ for
	color/leptocolor $SU(3)_C\times SU(3)_{\ell C}$
	confining gauge symmetry. 
	Standard model leptons, equipped with inherited
	Yukawa couplings, make then their appearance at
	the intermediate (gauge protected)
	$SU(3)_C\times SU(3)_L\times SU(3)_R$
	trinification stage as three pre-lepton composites.
	The addition of an exclusively prescribed
	(on anomaly free and Pauli exclusion grounds)
	spontaneously broken $SU(3)_F$ horizontal link
	to the unification chain, in the spirit of "One three
	to rule them all", strikingly preserves the anti-symmetric
	structure of the single family fermionic wave function.
	The threefold quark/lepton (spectators + composites)
	flavor chiral representation is then necessarily
	supplemented by a trinification singlet composite
	Majorana neutrino.
	The scheme serendipitously predicts a novel
	anomaly free lepto/dark portal.}

\begin{document} 
\maketitle
\flushbottom

\section{Introduction}
\label{sec:intro}

A long time ago, in what seems to be
another era, nucleons (protons and neutrons)
and electrons were considered indivisible
elementary particles.
Primarily owing to their discretely quantized
electric charges (Millikan integer sequence),
they govern atomic physics, and consequently
the structure of matter at the Bohr scale and
beyond.
Quarks have entered the game through the
group theoretical "eightfold way" path, and their
strong  local $SU(3)_C$ color interactions
have been proposed and experimentally
confirmed a bit later.
The integer charged baryons and mesons have
eventually become composites of fractionally
charged confined quarks.

The leptons, on the other hand, have fully preserved
their fundamental point-like perception during the
quark revolution.
Moreover, together with quarks they share the
standard $SU(3)_C \times SU(2)_L \times U(1)_Y$
electro/nuclear model, and in particular, contribute
their part to cancel the otherwise harmful ABJ
anomalies.
The latter feature takes us beyond the standard
model, crossing the gauge desert all the way to
grand unification.
While staying essentially the same for $SU(5)$,
anomaly cancelation becomes automatic for $SO(10)$
and $E(6)$.

By consistently embedding $SU(3)_C$ within
a parent $SU(4)_{PS}$, the unity of quarks
and leptons has been formally demonstrated
in Pati-Salam's "Lepton number as the fourth
color" \cite{PS}.
Numerically speaking, $SU(4)_{PS}$ unitarity
has been simply translated into the fundamental
sum rule $\frac{1}{3}+\frac{1}{3}+\frac{1}{3}-1=0$.
While such an imaginative idea has forcefully
opened the door for the left-right symmetric
grand unification variants $SO(10)\subset E(6)$,
once the spontaneous symmetry breaking
$SU(4)_{PS}\rightarrow SU(3)_C \times U(1)_{B-L}$
takes place, quarks and leptons are not treated on
equal symmetric footing any more. 
After all, one can always build an integer out of
thirds, but cannot vice versa build a third out
of integers.

The idea of accompanying $SU(3)_C$ by extra
 $SU(3)$ factors (with or without an underlying
 discrete symmetry) is not new, with
 $E(6)$-embeddable \cite{E6a,E6b,E6c,E6d}
 trinification \cite{333a,333b,333c,333d,333e,
 333f,333g,333h,333i}
 \begin{equation}
	 G_{tri}=SU(3)_C\times SU(3)_L
	 \times SU(3)_R
	 \label{Gtri}
\end{equation}
being the prototype example.
Other examples include LR-symmetric chiral
color $SU(3)_{CL}\times SU(3)_{CR}$
\cite{chiralC1,chiralC2}, confining
hypercolor (technicolor) $SU(3)_{HC}$
of various sorts (notably the Rishon model
\cite{Rishon1,Rishon2} and some of its derivatives
\cite{HCcomp1,HCcomp2,HCcomp3}),
spontaneously broken flavor
$SU(3)_F$ to account for the fermion threefold
family replication \cite{3F1,3F2}, and even leptocolor
$SU(3)_{\ell C}$ aimed to enhance
quark/lepton symmetry \cite{qlsym1,qlsym2,qlsym3,qlsym4,qlsym5,qlsym6}.
 On symmetry grounds, with "lepton number
as the fourth color" in mind, there is no group
theoretical reason why not considering "baryon
number as the fourth leptocolor" as well.
This would bring nucleons and electrons,
which anyhow share integer electric charges,
back to the same level of elementariness
(actually compositeness now), as used to be.
Such an idea can be minimally realized within
the framework of local
$SU(3)_C\times SU(3)_{\ell C}$ symmetry.
In fact, the latter symmetry has already been
introduced in the literature
\cite{qlsym1,qlsym2,qlsym3,qlsym4,qlsym5,qlsym6}.
However, the orthodox role of $SU(3)_{\ell C}$
as a spontaneously broken symmetry at heavy
mass scale is herby replaced by a totally different
scenario.
Namely, an unbroken symmetry, characterized
by a leptocolor scale substantially larger than
the ordinary color scale
$\Lambda_{\ell C}\gg\Lambda_C$
(with a gauge coupling constant
$g_{\ell C}\gtrsim g_C$ at $\Lambda_{\ell C}$).
Leptocolor is expected to enter its short
range confining mode much earlier than
$SU(3)_C$, at the stage where trinification
prevails.
As long as it stays unbroken, the latter local
symmetry is capable of gauge protecting a
flavor-chiral combination of spectator quarks
and composite leptons.

Composite model constructions are heavily
restricted by 't Hooft's anomaly matching equations
\cite{tHooft1,tHooft2,tHooft3,tHooft4}.
General solutions are hard to find, and the
more so solutions with integer coefficients
(even in cases where the restrictive
Apellquist-Carazzone decoupling condition
\cite{AC} is not applicable).
In fact, 't Hooft himself has pointed out that
$QCD$ with more than two flavors is not
natural, and made the interesting remark that to
construct tenable models with complete naturalness
for elementary particles one may need more types
of confining gauge theories besides $QCD$
(an example is provided in \cite{Ter}).
At any rate, contrary to 't Hooft's $3$-quark
model, where the associated
$SU(3)_L\times SU(3)_R \times U(1)_V$ flavor
symmetry is notably global, the latter symmetry
is almost fully gauged in our case, except for the
vector $U(1)_V$ remnant.
Unfortunately, our model is not natural in
't Hooft sense, with the problematic matching
equations being of course those involving
$U(1)_V$.
However, within the context of a flavor-chiral
model (unlike QCD), it makes it a bit easier for
us to assume that fermion masslessness is
protected by anomaly free gauge invariance.
While chiral symmetry suffices for this purpose,
anomaly cancellation is stronger, and furthermore
serves as a field theoretical consistency tool. 

The threefold family structure cannot and does
not stay out of the game.
It integrates in naturally and inductively by
means of adding another local $SU(3)_F$ link
to the $SU(3)$ chain
\cite{3chain1,3chain2,3chain3,3chain4,qlsym5,
3chain5,3chain6,3chain7}.
And quite remarkably, it does it without
upsetting the overall Pauli anti-symmetric
structure of the lepton wave function, a
phenomenon which we have already
encountered at the single family level.

\section{Composite leptons, spectator quarks}
\label{sec:comp/spec}

Let our $q\leftrightarrow \bar{\lambda}$ symmetric,
anomaly free, flavor-chiral representation of left handed
fermions transform as
\begin{equation}
	\psi_L=
	\begin{array}{|c ||c|c||c|c|}
	\hline ~ ~ & C & \ell  C & L & R\\ \hline
	\hline ~q_L~ & 3 & 1 & ~3^\star & 1\\
	\hline ~\bar{q}_L~ & ~3^\star & 1 & 1 & 3\\
	\hline ~\lambda_L~ & 1 & ~3^\star & 3 & 1\\
	\hline ~\bar{\lambda}_L~ & 1 & 3 & 1 & ~3^\star\\
	\hline \end{array}
\end{equation}
under the local quartification gauge group
\begin{equation}
	G_{quad}=[SU(3)_C\times SU(3)_{\ell C}]
	\times [SU(3)_L\times SU(3)_{R}] ~.
	\label{Gquad}
\end{equation}
Note that $q \leftrightarrow \bar{\lambda}$ under
$\{ C \leftrightarrow \ell C,L \leftrightarrow R\}$,
so that one cannot really tell quarks from pre-leptons
at this stage.
The representation is $C\ell C$ and $LR$ real,
separately, yet overall complex, forming a moose chain.
Conspicuous by its absence is the $(1,1;3,3^\star$)
leptonic piece which has been bifurcated into the
$\lambda+\bar{\lambda}$ pair. 
Starting with $G_{quad}$, we end up
with unbroken
$SU(3)_C\times SU(3)_{\ell C}\times U(1)_Q$
once the dust settles down. 
The electric charge reads
\begin{equation}
	Q=T_{3L}+T_{3R}+\frac{1}{2}
	\left(Y_L+Y_R \right) ~,
\end{equation}
so that $Q[q]+Q[\lambda]$=0.
At any rate, we keep the option of inductively
enlarging $G_{quad}$, if needed, when proceeding
beyond the scope of the single family level.

In analogy with the color singlet baryons which are
made of three quarks, we now attempt to construct
leptocolor singlet composites using
\begin{equation}
	\lambda_L (1,3^\star;3,1)~,\quad
	\lambda_R (1,3^\star;1,3)
\end{equation}
as our building blocks.
Here, invoking the traditional basis, $f_R$ denotes
the right handed  Weyl partner of the left handed
$\bar{f}_L$
(they transform into each other under charge conjugation). 
In turn, $\lambda_L$ and $\lambda_R$ transform alike,
that is $\sim 3^\star$ under $S(3)_{\ell C}$.
We are after a set of left handed composites which
would compensate for the $SU(3)_{L,R}$ triangle
anomalies contributed by the quark spectators. 
The fact that $\lambda_R \lambda_R \lambda_R$ and
$\lambda_L \lambda_L \lambda_R$ are right-handed
leaves us with the two left-handed candidates: 

\medskip\noindent (i) $\lambda_L \lambda_L \lambda_L$: 
\quad
From $SU(2)_L\times SU(2)_R^\dagger$
Lorentzian point of view, we can have $(2+2+4,1)$,
spinors nicely included (in fact, there are two of them).
But this turns out irrelevant given that under local
$SU(3)_L\times SU(3)_R$, the resulting configuration
$(1+8+8+10,1)$ does not contain triplets nor
anti-triplets.
Put it in other words, no standard model leptonic
candidates are available.
Note in passing that, unlike complex $(10,1)$, the
real representations $(1+8+8,1)$ cannot anyhow
be protected by gauge invariance.
The $(1,1)$ piece will play a role in a later stage.

\medskip\noindent (ii) $\lambda_L \lambda_R \lambda_R$:
\quad The Lorentz representation $(2,1+3)$ also
contains the mandatory $(2,1)$ piece, but unlike in
the previous case, one faces now $(3,3^\star+6)$
under $SU(3)_L\times SU(3)_R$, including the
mandatory $(3,3^\star)$ representation.
With this in hands, we can now complete our composite
level lepton identification, namely
\begin{equation}
	\ell_L +\bar{\ell}_L=(1,1;3,3^\star) ~.
\end{equation}
Together with the quarks and anti quarks, hereby treated
as spectators, the leptocolor singlets close upon
\begin{equation}
	\psi_L^{tri}=
	\begin{array}{|c ||c|c|c|}
	\hline ~ & C & L & R \\ \hline
	\hline ~q_L~ & 3 & ~3^\star & 1\\
	\hline ~\bar{q}_L~ & ~3^\star & 1 & 3\\
	\hline ~\ell_L+\bar{\ell}_L~
	& 1 & 3 & ~3^\star \\	\hline \end{array}
\end{equation}
forming a complex single family anomaly free,
that is $3\times (-1)+9\times 0+3 \times 1=0$, 
trinification representation.
It is important to note that the $SU(3)_R$
triangle anomaly $A(6)=-7A(3^\star)$ makes
the $(1,1;3,6)$ representation irrelevant
for our anomaly cancelation purposes.
Had we incorporated $(1,1;3,6)$, then
owing to $A_{3L}+A_{3R}=3(2+7)\neq 0$,
there would
be no way to simultaneously cancel both
$SU(3)_{L,R}^3$  anomalies $A_{3L,3R}$,
respectively.
As long as $G_{tri}$ stays unbroken, this
complex set is protected and remains massless.
Other composite states, such as (say)
$\lambda_L \lambda_L \lambda_L$ which
belongs to $SU(3)_L / Z_3$, are not
protected from picking up the large
$\Lambda_{\ell C}$ mass scale.

\begin{figure}[h]
	\centering
	\includegraphics[scale=0.28]{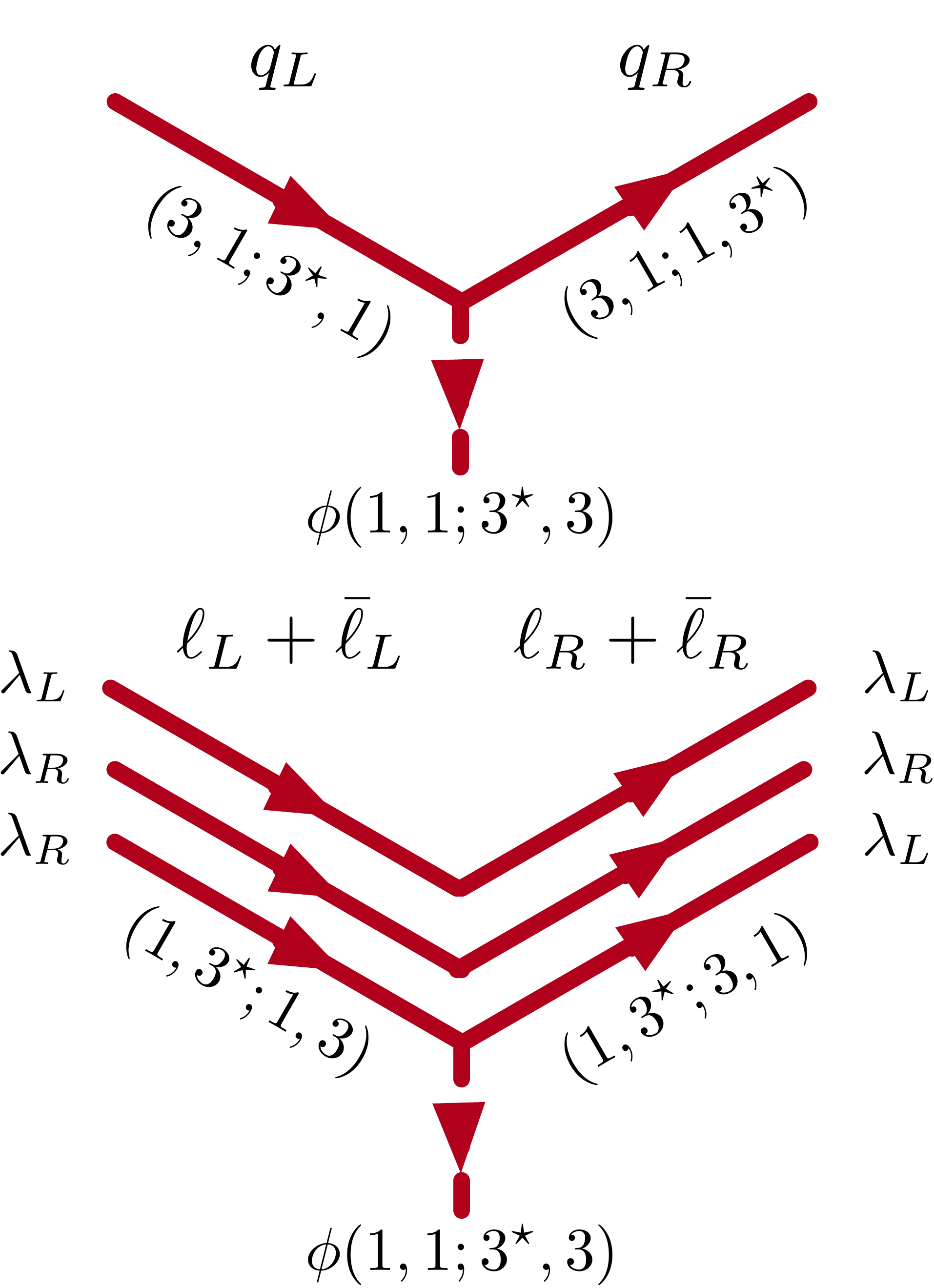}
	\caption{\label{Y}
	Yukawa at trinification:
	Spectator quarks and composite leptons
	share similar type Yukawa diagrams, involving
	the one and the same Higgs scalar.
	Only one pre-lepton is actively involved,
	while two others stay inertly out of the game.}
\end{figure}

\begin{figure}[h]
	\centering
	\includegraphics[scale=0.28]{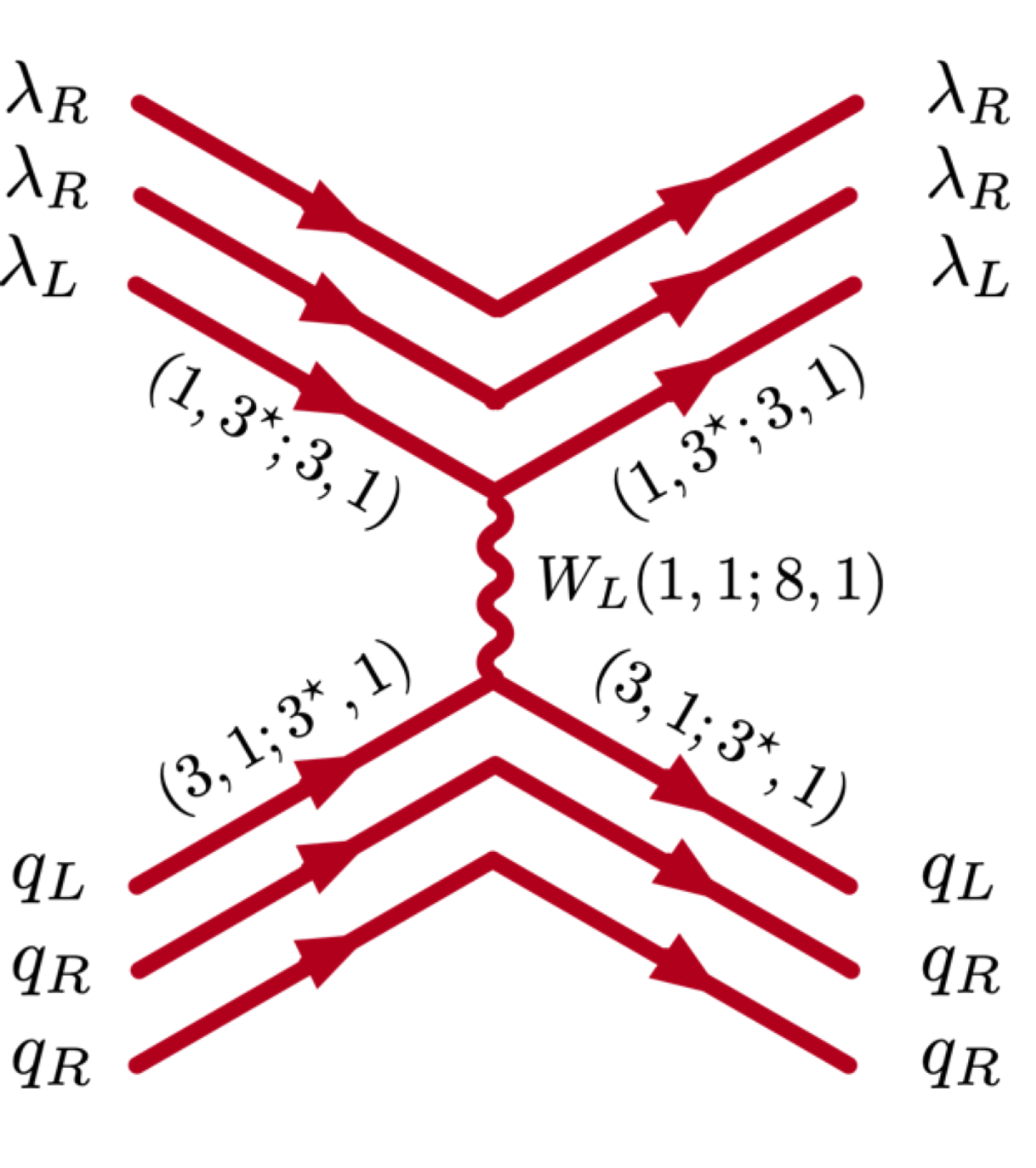}
	\caption{\label{Beta}
	Beta decay as depicted at the fundamental 
	quark/pre-lepton level:
	The weak gauge boson mediators $W_L^{\pm}$
	are members of the $SU(3)_L$ adjoint
	representation.}
\end{figure}

The Yukawa interactions, introduced at the
fundamental Lagrangian level, forcefully reappear
at the intermediate trinification stage.
As depicted in Fig.(\ref{Y}), spectator quarks
and composite leptons share the one and the
same Yukawa coupling diagrams.
As far as the leptons are concerned, only one
pre-lepton is involved in the coupling, while its
two companions stay inert.
A similar situation occurs in the leading order
of beta decay Fig.(\ref{Beta}).

\section{Threefold family structure}
\label{sec:3fold}

Once the single quark/lepton family model
has been established, the obvious challenge is
to incorporate the underlying threefold family
structure, and do it naturally without any
superfluous replication.
With the underlying motto of "One three
to rule all flavors", we are driven to the quite
traditional step of introducing an extra local
$SU(3)_F$ group factor ($F$ stands for Family).
One has to be careful though, especially when
dealing with composite models, to ensure the
overall anti-symmetry of the lepton wave function
under the interchange of the two identical
$\lambda_R$s.
This is however not automatic and should be
regarded as a consistency test for the model
under construction.

Without loosing generality, let
$\lambda_L\sim 3^\star$ under $SU(3)_F$,
so that
\begin{equation}
	\lambda_L \lambda_L \lambda_L
	\sim 3^\star \times 3^\star \times 3^\star
	=1+8+8+10^\star ~.
	\label{LLL}
\end{equation}
There are still two possibilities
regarding the assignments of $\bar{\lambda}_L$
(or equivalently of $\lambda_R$).
To choose between these two, one needs
to first check the associated transformation
laws: 

\medskip\noindent (i)
$\bar{\lambda}_L\sim 3$ ($\lambda_R\sim 3^\star$)
implies
\begin{equation}
	\lambda_L \lambda_R \lambda_R
	\sim 3^\star \times 3^\star \times 3^\star
	=1+8+8+10^\star ~.
\end{equation}

\noindent (ii)
$\bar{\lambda}_L\sim 3^\star$ ($\lambda_R\sim 3$)
implies
\begin{equation}
	\lambda_L \lambda_R \lambda_R
	\sim 3^\star \times 3 \times 3
	=3+3+6^\star+15 ~.
	\label{LRR}
\end{equation}
And since the left handed composites must include
a $3$ (or a $3^\star$) to consistently enter the
family picture, it is only the second option which
can deliver.
The main bonus is fixing the inner structure of
the tenable composite fermion, namely choosing
$\lambda_L \lambda_R \lambda_R$ rather than
$\lambda_L \lambda_L \lambda_L$.
It is by no means trivial that the very same conclusion
was drawn previously, where only
$\lambda_L \lambda_R \lambda_R$ was shown to
transform as required, that is via $(3, 3^\star)$, under
the electro/weak $SU(3)_L\times SU(3)_R$.
Still, see eq.(\ref{LRR}), there are two
$\lambda_L \lambda_R \lambda_R$ triplet candidates.
Using $SU(3)_F$ spinor indices, they are

\begin{subequations}
\begin{align}
	\text{anti-symmetric:}\quad
	\epsilon^{iqp}\epsilon_{pjk}\lambda_{Lq}
	\lambda_R^j \lambda_R^k ~,~
	\label{asym}\\
	\text{symmetric:}\quad
	\lambda_{Lp}(\lambda_R^i \lambda_R^p
	+\lambda_R^p \lambda_R^i) ~.
	\label{sym}
\end{align}
\end{subequations}
The choice between these two candidates will be
eventually made on symmetry (or anti-symmetry)
grounds when the full picture will be in front of us.

A word of caution is now in order.
Note that $\lambda_L \lambda_R \lambda_R$
is expected to transform according to the
$(\frac{1}{2},0)$ representation under Lorentz
group $SU(2)_L\times SU(2)_R^\dagger$. 
This means that $\lambda_R \lambda_R$ must
be treated as a Lorentz scalar, that is
$\epsilon_{ij}\lambda_R^i \lambda_R^j$
(here $i,j$ are $SU(2)_R$ spinor indices).
To sharpen this point, with the focus on its underlying
antisymmetric Lorentz structure, a somewhat better
notation for the composite is perhaps
$\lambda_L [\lambda_R \lambda_R]$.
The same should hold of course for the left handed
composite fermions $\lambda_L \lambda_L \lambda_L$
as well, whose inner Lorentz structure is better
manifested via $\lambda_L [\lambda_L \lambda_L]$.

At any rate, recalling that $\lambda_L$ and
$\bar{\lambda}_L$ are both $SU(3)_F$ anti-triplets,
there seems to be just one way to cancel the potentially
newly induced ABJ anomalies.
Quarks and anti-quarks are called to the rescue,
and they do it quite naturally by means of
\begin{equation}
	\lambda_L, \bar{\lambda}_L \sim 3^\star
	\quad \Longrightarrow\quad
	q_L, \bar{q}_L \sim 3 ~,
\end{equation}
to be regarded a fundamental feature of the
quark/lepton symmetry.

While $\lambda_L, \bar{\lambda}_L \sim 3^\star$,
notice that the emerging composite lepton state
$\lambda_L \lambda_R \lambda_R$, previously identified
as $\ell_L+\bar{\ell}_L$, transforms under $SU(3)_F$
precisely like $q_L, \bar{q}_L $, that is
$\lambda_L \lambda_R \lambda_R \sim 3$.
In turn, with the addition of $SU(3)_F$, the attractive
trinification scheme, dynamically encountered at the
single family level following leptoquark confinement, is
apparently in jeopardy.
The accumulated anomaly of twenty seven $SU(3)_F$
triplets must be taken care of in order to gauge protect
the masslessness of the three
$\{q_L, \bar{q}_L, \ell_L+\bar{\ell}_L\}$ flavor-chiral standard
trinification families.
One is thus after a leptocolor singlet, $SU(3)_F$ non-trivial
composite fermion which, on self consistency grounds,
must further transform as a total multi-singlet $(1,1,1)$
under $SU(3)_C \times SU(3)_L \times SU(3)_R$.

\begin{figure}[h]
	\centering
	\includegraphics[scale=0.28]{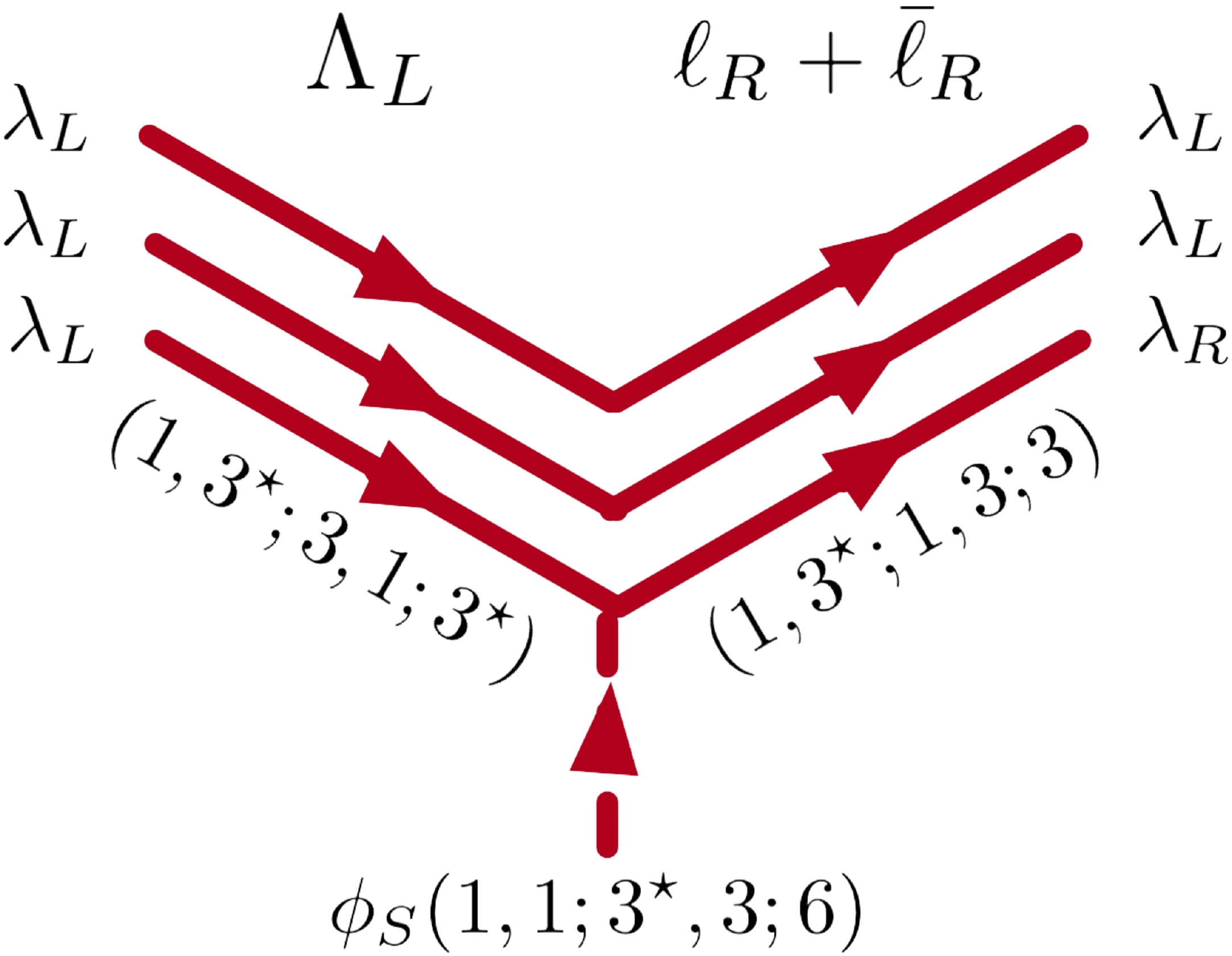}
	\caption{\label{newY} Lepto/Dark portal:
	$\phi\ell \Lambda$ Yukawa coupling involves
	the same Higgs scalar which couples (and
	generates masses upon developing VEV) to
	standard model fermions. 
	$\phi_S \sim 6$ under $SU(3)_F$ couples
	symmetrically in family space.}
\end{figure}

\section{Lepto/Dark portal}
\label{sec:portal}

It comes with no surprise that precisely such
a composite fermion does exist in the model
and is at our service (see anomaly details below).
We refer of course to
$\Lambda_L\equiv\lambda_L \lambda_L
\lambda_L \sim 10^\star$ which has already made
 its appearance in Eq.(\ref{LLL}).
The fact that $\Lambda_L$ is totally symmetric in
its $SU(3)_F$ spinor indices is crucial.
By the same symmetry token, it chooses for us the
symmetric $\lambda_L \lambda_R\lambda_R$ triplet
candidate Eq.(\ref{sym}).
The
anti-symmetric structure of the lepton wave function
under identical pre-lepton interchange, encountered
already at the single family model, is preserved (it
could not have been done differently) in the modified
threefold family version.

In 4-dim, the $SU(n)$ triangle anomaly table
\cite{anomaly1,anomaly2,anomaly3} offers the relation
\begin{equation}
	A(\square\square\square)=
	\frac{1}{2}(n+6)(n+3)A(\square) ~,
	\label{A}
\end{equation}
where the box stands for the fundamental
representation, and in general we have
$A(r^\star)=-A(r)$.
In particular for $n=3$, the above formula reads
\begin{equation}
	A(10^\star)+27 A(3)=0 ~,
\end{equation}
thereby elevating
$\Lambda_L=\lambda_L \lambda_L \lambda_L$
to the level of a mandatory supplement to the
threefold quark/lepton family (spectators $+$
composites) flavor chiral representation.
Altogether, the latter is given explicitly by
\begin{equation}
	\psi_L^{tri}=
	\begin{array}{|c ||c|c|c||c|}
	\hline ~ & C & L & R & F\\ \hline
	\hline ~q_L~ & 3 & ~3^\star & 1 & 3\\
	\hline ~\bar{q}_L~ & ~3^\star & 1 & 3 & 3\\
	\hline ~\ell_L+\bar{\ell}_L \equiv
	\lambda_L \lambda_R \lambda_R~
	& 1 & 3 & ~3^\star & 3 \\
	\hline ~\Lambda_L \equiv
	\lambda_L \lambda_L \lambda_L
	~ & 1 & 1 & 1 & ~10^\star\\
	\hline \end{array}
\end{equation}
The overall anomaly is now
$9\times 1+9\times 1+9\times 1-1\times 27=0$. 
Had we started from $SU(n)_F$, thereby facing
the anomaly equation $3\times 9\times 1
-1\times \frac{1}{2}(n+6)(n+3)=0$,
the special case of $n=3$ would have been
singled out on anomaly free and Pauli exclusion
grounds.
This establishes a novel group theoretical
interplay between the single fermionic family
and its threefold replication.

\begin{figure}[h]
	\centering
	\includegraphics[scale=0.28]{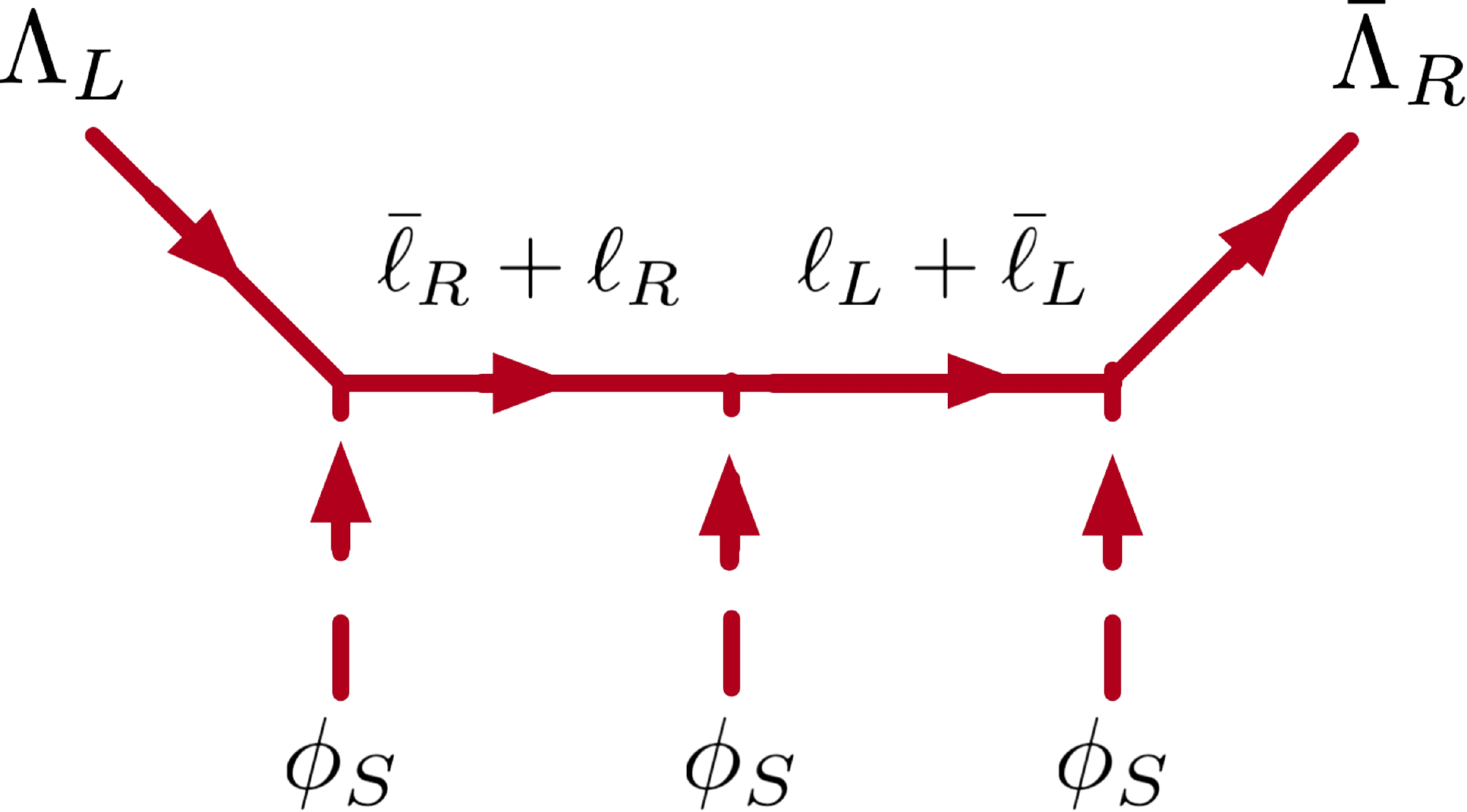}
	\caption{\label{Majorana} A possible dark mass origin:
	Depicted is a tree level diagram, involving exactly
	three Yukawa couplings, by which the trinification
	singlet fermion $\Lambda_L$ acquires its Majorana
	mass.}
\end{figure}

It is only when $SU(3)_F$ gets broken down,
and leaves no residual gauge symmetry to protect
$\Lambda_L$ from acquiring the heavy mass scale
involved, that one finally faces the familiar
$E(6)$-unifiable trinification scheme (threefold
'superfluous' replication included).
Be reminded that Lorentz invariance allows for
two types of mass terms.
Dirac mass term ${f_{1L}}^{\dagger}f_{2R}$
involves two opposite helicity Weyl fermions
$f_{1L}$ and $f_{2R}$ which must then carry
the same conserved charges
$Q(f_{1L})-Q(f_{2R})=0$.
Majorana mass term, say $f_{1L}f_{2L}$, involves
two (may be the same) same helicity Weyl fermions,
with $Q(f_{1L})+Q(f_{2L})=0$.
By acquiring a heavy Majorana mass
$m_{\Lambda}\Lambda_L \Lambda_L$ at
this level (no $\Lambda_R$ at our disposal to
produce a Dirac mass as well), $\Lambda_L$ plays
here a role similar to that of the $\nu_R$ in
the standard model.
In fact, appreciating its electro/nuclear neutrality,
it may trigger the very same neutrino seesaw
mechanism provided it can couple to standard
model particles.
A possible (there may be others) tree level
mechanism by which $\Lambda_L$ can acquire
its Majorana mass is suggested in Fig.(\ref{Majorana}).
Owing to the $SU(3)_F$ multiplication feature
$10 \times 10 \subset 6 \times 6 \times 6$,
it involves exactly three Yukawa couplings.
The relevant inner fermion lines are 
associated with heavy Majorana neutrinos 
(there are two per trinification family)
which live outside minimal $G_{LR}=SU(2)_L
\times SU(2)_R\times U(1)_{L}\times U(1)_{R}$.
Note that $\Lambda_L$ Yukawa mix with both heavy
LR components $(2_1,2_{-1})$ and $(1_{-2},1_2)$
of $\ell_L+\bar{\ell}_L$.
Two of the three scalar lines end with
a $\langle 2_{-1},2_1 \rangle$ VEV, while the third
one ends with a $\langle(1_2,1_{-2}) \rangle$  VEV, 
respectively.

To refresh the Higgs mechanism scheme in the
presence of the newly introduced $SU(3)_F$
group, we first observe that the Yukawa interactions
are now governed by $\phi (1,1;3^\star,3;3^\star+6)$.
They couple symmetrically ($\phi_S \sim 6$) and/or
anti-symmetrically ($\phi_A \sim 3^\star$) in the
family space.
Given the odd number of families, $\phi_A$ is not
allowed to Yukawa couple by itself, as otherwise one
family stays massless and the other two degenerate. 
This makes the symmetric Yukawa $\phi_S$ couplings
essential.
The novel ingredient is that the ordinary Yukawa
couplings, depicted in Fig.(\ref{Y}), are now
supplemented, as is demonstrated in Fig.(\ref{newY}), by
$\phi_S(\ell_R+\bar{\ell}_R)\Lambda_L$.
This can be interpreted as a lepto/dark portal.
After all, $\Lambda_L$ is a Majorana heavy trinification
singlet, striped from all electro/nuclear interactions, 
capable of seesaw mixing
$(\theta \sim m_\ell/ m_{\Lambda})$
with standard model fermions.

\section{Epilogue}
\label{sec:Epilogue}

While quarks and leptons are equal rights
members of the standard electro/nuclear
theory, they are not really treated on
symmetrical footing.
Lepton number can serve as the fourth
"color", but it is harder to imagine baryon
number serving as the fourth "leptocolor".
The fact that an integer can be built out
of thirds, but not vice versa, naively suggests
that leptons (integer charges) are made
out of three pre-leptons (fractional charges),
such that one would not be able to tell
quarks from pre-leptons at the fundamental
level.
A group theoretical realization of this idea
is provided in this paper, with $E(6)$-unifiable
trinification emerging in the intermediate stage,
and with the threefold family structure inductively
incorporated.
Thus, to answer the question in the title in a
challenging (not to say provocative) way, leptons
need not be, at least theoretically, as elementary
as quarks.

A few comments are finally in order:
\newline (i) In analogy with LR symmetry, one
may anticipate $C \ell C$ symmetry to be
spontaneously violated, such that leptocolor
becomes the first interaction to confine.
Until such a symmetry breaking mechanism is
introduced, one has to explicitly assume
$g_{\ell C}\gtrsim g_C$.
\newline (ii) We can say nothing compelling at
this stage about the quark/lepton mass spectrum
and mixings.
While we rely on the dynamical emergence of
the intermediate trinification phase, we have not
incorporated any particular trinification model
(from existing ones \cite{333a,333b,333c,333d,333e,
 333f,333g,333h,333i} or beyond) for probing fine
 details.
\newline (iii) It is not clear what ingredients can be
added to composite models in general, and to our
model in particular, in order to achieve (or at least
tenably avoid) the restrictive 't Hooft naturalness
requirement.
\newline (iv) In addition, we have not investigated
at this stage what could be the benefits, if at all, of
incorporating supersymmetry. We leave it for a
future investigation.

\acknowledgments
I would like to dedicate this paper to the memory
of Prof. Kameshwar C. Wali (Syracuse University). 
A scholar, a mentor, a colleague and a very dear friend.

\end{document}